\documentclass[twocolumn,showpacs,preprintnumbers,amsmath,amssymb]{revtex4-1}

\usepackage{graphicx}% Include figure files
\usepackage{dcolumn}% Align table columns on decimal point
\usepackage{bm}% bold math
\usepackage{epsfig}

\begin{document}

%\preprint{hep/ph}

\title{Higher moments of multiplicity fluctuations in a hadron-resonance gas with exact conservation laws}

\author{Jing-Hua Fu}
\email{fujh@mail.ccnu.edu.cn} \affiliation{Institute of Particle
Physics, Central China Normal University, Wuhan 430079, PR China \\
Key Laboratory of Quark \& Lepton Physics (CCNU), Ministry of
Education, PR China}
\date{\today}

\begin{abstract}
Higher moments of multiplicity fluctuations of hadrons produced in central nucleus-nucleus collisions are studied within the
hadron-resonance gas model in the canonical ensemble. Exact conservation of three charges, baryon number, electric charge,
and strangeness, is enforced in the large volume limit. Moments up to the forth order of various particles are calculated at SPS, 
RHIC and LHC energies. 
The asymptotic fluctuations within a simplified model with only one conserved charge in the canonical ensemble are discussed where simple analytical expressions for moments of multiplicity distributions can be obtained. Moments products of net-proton, net-kaon, and net-charge distributions in Au + Au collisions at RHIC energies are calculated. The pseudo-rapidity coverage dependence of net-charge fluctuation is discussed.
\end{abstract}

% PACS, the Physics and Astronomy
                             % Classification Scheme.
%\keywords{Suggested keywords}%Use showkeys class option if keyword
                              %display desired
\maketitle

\section{Introduction}
%\label{}
The Beam Energy Scan (BES) program at the Relativistic
Heavy-Ion Collider (RHIC) facility aims at studying in detail
the QCD phase structure. 
Nonmonotonic behavior in the fluctuations of
globally conserved quantities, such
as net-baryon, net-charge, and net-strangeness numbers,
as a function of beam energy are believed to be good 
signatures of a phase transition and a QCD critical point (CP)
\cite{AsakawaHeinzMullerPRL2000, Karsch2011plb}. 

Event-by-event distributions of
conserved quantities within a limited acceptance are
characterized by the moments, such as the mean (M), the
standard deviation ($\sigma$), the skewness (S),  and the kurtosis ($\kappa$). 
These moments are related to the corresponding higher-order thermodynamic
susceptibilities and to the correlation length of the system, with increasing sensitivity for higher order moments \cite{Stephanov99prd,Stephanov09prl}. The products
of the moments, such as $\sigma^2/M$, $S\sigma$, and $\kappa\sigma^2$, are constructed
in order to cancel the volume term and allow for a direct comparison of experimental measurement and theoretical calculation. 

Particle production at the time of
freeze-out in heavy-ion collisions from SIS up to LHC energies
exhibit thermal characteristics and are well described by the hadron-resonance gas (HRG) model \cite{PBM06npa}. Since the
sensitivity to critical dynamics grows with the increasing order,
the values of higher-order moments of charge fluctuations can
differ significantly from the results of HRG along the freeze-out
curve even if lower-order moments agree \cite{Karsch2011plb}. The
HRG model results on moments of charge fluctuations can serve as a
theoretical baseline for the analysis of heavy-ion collisions to properly assess the discriminating power of these observables.

In the applications of the HRG model in nucleus-nucleus collisions, the grand canonical ensemble (GCE) is most commonly used ensemble because it is the most convenient one from a technical point of view, and because of the fact that both the canonical
ensemble (CE) and microcanonical ensemble (MCE) become equivalent to the GCE in the thermodynamic limit. 

The applicability of the statistical approach formulated within various statistical ensembles has been considered for average quantities.  In the GCE, chemical potentials and temperature are introduced to control the material and motional conservation on the average, which implies fluctuations around the average from event to event. The GCE can be used when the number of carries of conserved charge is large enough and the fluctuations can be neglected \cite{PBM2003QGP}. 
The CE, where the material conservation laws are strictly obeyed, is relevant for systems with a large number of all produced particles, but a small number of carriers of conserved charges like electric charge, baryon number, strangeness or charm \cite{HagedornRedlich1985ZPhysC,CleymansKoch1991ZPhysC}. This may happen not only in elementary $pp$, $p\bar{p}$ and $e^+e^-$ collisions \cite{BecattiniHeinz1997ZPhysC}, but also in $pA$ or even $AA$ collisions \cite{CleymansRedlich1991ZPhysC,CleymansCanonical1997prc,PBMNetProton2011prc}.  Finally, the MCE, where both the motional and material conservation laws are strictly fulfilled, has been used for elementary collisions at low energies where a small number of particles is produced \cite{Werener1995PRC}. 

The different statistical ensembles are not equivalent for small systems. When the system volume increases, the average quantities obtained with the GCE, CE, and MCE approach each other in the large-volume limit, $V\rightarrow\infty$, which is called
the thermodynamical equivalence of the statistical ensembles. 
The thermodynamic equivalence, however, applies only to the average quantities obtained in different ensembles, but is not true for the scaled variance of particle number fluctuations \cite{ConservationGorensteinPRC2004, ConservationGorensteinPRC2005,ConservationBecattiniPRC2005}. There is a qualitative difference in the properties of the mean multiplicity and the scaled variance of multiplicity distribution in statistical models.
Expressed in terms of the scaled variance, the particle number fluctuations have been found to be suppressed in the CE and MCE comparing to the GCE and this suppression survives in the limit $V\rightarrow\infty$. The differences of the scaled variance from different ensembles preserved in the thermodynamic limit. 
The scaled variance is sensitive to conservation laws obeyed
by a statistical system. 

Most calculations \cite{Jinghua2009PLB,Jinghua2013PLB,Subhasis2014PRC,Nahrgang2014PLB,Nahrgang2015EPJC,Mohanty2016PRC,KarschRedlich2016PRC} on higher moments of multiplicity fluctuations in the HRG model employed only the GCE treatment.
To approximately achieve conditions of the GCE, one is required to study fluctuations in a restricted phase space \cite{Koch2013PRC, Nahrgang2012EPJC}. This is usually fulfilled since there are cuts in rapidity and/or transverse momentum of the detected particles. The smaller the fraction of observed particles the smaller is the effect of global charge conservation.  However, care has to be taken that these cuts do not destroy the underlying correlations related to the physics one tries to access. 

Measurements on higher moments of net-proton, net-kaon and net-charge multiplicity distributions have been recently carried out by the STAR and PHENIX Collaboration 
in Au + Au collisions at several collision energies and centralities
\cite{STARNetProtonPRL2014,STARNetChargePRL2014,PHENIXNetChargePRC2016,LuoXFCPOD2014,JochenQM2015}. No significant deviation from the Poisson expectation is observed within uncertainties for net-charge and net-kaon moments products. Some pronounced structure is found in the energy dependence of $\kappa\sigma^2$ of net-proton distributions from the $0\%-5\%$ most central collisions within $0.4<p_T<2$ GeV/$c$ and at mid-rapidity $|y|<0.5$ for energies below 39 GeV. This same measurement becomes close to the Poisson expectations when the $p_T$ or rapidity acceptance decreases. 

As the collision energy decreases and phase space coverage increases, a larger fraction of the total (conserved) charge is observed, and the impact of conservation laws on the multiplicity and charge fluctuations might show up. 
The higher (third and forth) order moments of multiplicity distributions in the HRG model have not been calculated explicitly in the CE. The present work present a first calculation of the higher order fluctuations in the general multispecies hadron gas including all mesons up to the
$K_4^*(2045)$, baryons up to the $\Omega^-$ and carrying three additive charges, baryon number, strangeness, and electric charge, in the CE.   

This paper is organized as follows: In the next section the partition function and the first four moments of multiplicity distributions in the CE are presented. The complete expression for high order moments of multiplicity distributions in the CE are complicated. To get a more direct idea about the extra-Poisson fluctuations in the CE,
a simple illustrative example with only one type of conserved charge is discussed in Sec. \ref{SecAsymptotic}, where simple analytical expressions for moments of multiplicity distribution can be obtained. In Sec. \ref{SecFreezeout} and \ref{SecResonance} the Monte Carlo simulation procedures and  resonance decay contributions are described.   
In Sec. \ref{SetNetparticle} the first four moments of net-proton, net-kaon and net-charge multiplicity distributions are calculated in the HRG model in the CE with exact conservation of baryon number, electric charge and strangeness. The CE results are compared with the GCE results. As an application of the model results, the pseudo-rapidity coverage dependence of net-charge fluctuation is discussed in Sec. \ref{SecRapCover}.  

\section{Multiplicity fluctuations in the canonical ensemble}
\label{SecCanonical}

In the HRG model the partition function contains all relevant
degrees of freedom of the confined, strongly interacting matter and
implicitly includes interactions that result in resonance formation.
The GCE partition function of a hadron-resonance gas can be written as a product of partition functions of all hadrons and resonances, following the notations of Ref.  \cite{ConservationBecattiniPRC2005},   
\begin{equation}
	Z_{\mathrm{GC}} (\{\lambda_j\})= 
	\prod_j \exp \left[ \sum_{n_j=1}^\infty \frac{{z_{j(n_j)}} \lambda_j^{n_j}}{n_j} \right],
\end{equation}
where
\begin{equation}\label{zjn}
	{z_{j(n_j)}} = (\mp 1)^{n_j+1}\frac{g_j V}{2\pi^2 n_j}T m_j^2{\rm K}_2\left(\frac{n_j m_j}{T}\right),
\end{equation}
$K_2$ is the modified Bessel function, $\lambda_j$ is the fugacity for each particle species $j$, $m_j$ is the hadron mass, $g_j = 2J_j+1$ is the spin degeneracy, $V$ is the volume of the hadron gas, the upper sign for fermions and the lower for bosons. Retaining just the first ($n_j = 1$) term corresponds to the classical Boltzmann approximation.

The  partition function in the CE does not factorize into one-species
expressions because of the constraint of fixed charges. Consider a hadron gas 
with three abelian charges, i.e., baryon number $B$, strangeness $S$, and electric charge 
$Q$. Denote them by a vector $\vec{Q} = (Q_1,Q_2,Q_3) = (B,S,Q)$  
whose components are these charges and by $\vec{q}_j = (q_{1,j},q_{2,j},q_{3,j}) = (b_j,s_j,q_j)$
the vector of charges of the  hadron species $j$.
The canonical partition function with charges $\vec{Q}$ can be written as 
\begin{equation} \label{projection}
	Z_{\vec{Q}} = \left[ \prod_{i=1}^{3} \frac{1}{2\pi} 
	\int_0^{2\pi}\mathrm{d}\phi_i \; e^{-i Q_i\phi_i}\right] Z_{\mathrm{GC}}(\{\lambda_j\}),
\end{equation}
where Wick-rotated fugacities $\lambda_j=\exp[i \sum_i q_{i,j} \phi_i]$ are 
introduced in the GCE partition function $Z_{\mathrm{GC}}$. 

The moments of multiplicity distributions of a set $h$ of hadron
species can be calculated by inserting a suitable fictitious fugacity into the function $Z_{\mathrm{GC}}$, i.e., replacing $\lambda_j$ with $\lambda_h\lambda_j$ in Eq. (\ref{projection}) if $j \in h$, and taking derivatives with respect to $\lambda_h$. The results of the first four moments are:
\begin{eqnarray} \label{CEmean}
	\langle N_h \rangle &=& \displaystyle{ 
		\frac{1}{Z_{\vec{Q}}} \frac{\partial Z_{\vec{Q}}}{\partial \lambda_h}
		\Bigg|_{\lambda_h=1}
	}
	=  \sum_{j\in h} 
	\sum_{n_j=1}^\infty  z_{j(n_j)} \frac{Z_{\vec{Q}-n_j\vec{q}_j}}{Z_{\vec{Q}}}, 
\end{eqnarray}
\begin{widetext}
\begin{eqnarray}\label{CEvariance}
\langle N_h^2 \rangle &=& \displaystyle{
	\frac{1}{Z_{\vec{Q}}} \left[\frac{\partial}{\partial\lambda_h}\left(\lambda_h
	\frac{\partial Z_{\vec{Q}}} {\partial \lambda_h}\right)\right]_{\lambda_h=1}
}\nonumber\\
&=& \sum_{j\in h}\sum_{n_j=1}^\infty n_j z_{j(n_j)} 
\frac{Z_{\vec{Q}-n_j\vec{q}_j}}{Z_{\vec{Q}}} + \sum_{j\in h} \sum_{n_j=1}^\infty 
z_{j(n_j)} \sum_{k \in h} \sum_{n_k=1}^\infty z_{k(n_k)} 
\frac{Z_{\vec{Q}-n_j\vec{q}_j-n_k\vec{q}_k}}{Z_{\vec{Q}}}, \\
	\label{CEcumulant3}
	\langle N_h^3 \rangle &=& \displaystyle{
		\frac{1}{Z_{\vec{Q}}} \left[\frac{\partial}{\partial\lambda_h}\left(\lambda_h\frac{\partial}{\partial\lambda_h}\left(\lambda_h
		\frac{\partial Z_{\vec{Q}}} {\partial \lambda_h}\right)\right)\right]_{\lambda_h=1}
	}\nonumber\\
	&=& \sum_{j\in h} \sum_{n_j=1}^\infty n^2_j z_{j(n_j)} 
	\frac{Z_{\vec{Q}-n_j\vec{q}_j}}{Z_{\vec{Q}}} + 3\left[\sum_{j\in h} \sum_{n_j=1}^\infty 
	n_jz_{j(n_j)} \sum_{k \in h} \sum_{n_k=1}^\infty z_{k(n_k)} 
	\frac{Z_{\vec{Q}-n_j\vec{q}_j-n_k\vec{q}_k}}{Z_{\vec{Q}}}\right]\nonumber \\
	& &+ \sum_{j\in h} \sum_{n_j=1}^\infty 
	z_{j(n_j)} \sum_{k \in h} \sum_{n_k=1}^\infty z_{k(n_k)}\sum_{l \in h} \sum_{n_l=1}^\infty z_{l(n_l)} 
	\frac{Z_{\vec{Q}-n_j\vec{q}_j-n_k\vec{q}_k-n_l\vec{q}_l}}{Z_{\vec{Q}}}, \\
\label{CEcumulant4}
	\langle N_h^4 \rangle &=& \displaystyle{
		\frac{1}{Z_{\vec{Q}}} \left[\frac{\partial}{\partial\lambda_h}\left(\lambda_h\frac{\partial}{\partial\lambda_h}\left(\lambda_h\frac{\partial}{\partial\lambda_h}\left(\lambda_h
		\frac{\partial Z_{\vec{Q}}} {\partial \lambda_h}\right)\right)\right)\right]_{\lambda_h=1}
	}\nonumber\\
	&=& \sum_{j\in h} \sum_{n_j=1}^\infty n^3_j z_{j(n_j)} 
	\frac{Z_{\vec{Q}-n_j\vec{q}_j}}{Z_{\vec{Q}}} + 4\left[\sum_{j\in h} \sum_{n_j=1}^\infty 
	n^2_jz_{j(n_j)} \sum_{k \in h} \sum_{n_k=1}^\infty z_{k(n_k)} 
	\frac{Z_{\vec{Q}-n_j\vec{q}_j-n_k\vec{q}_k}}{Z_{\vec{Q}}}\right]\nonumber \\
	& &+ 3\left[\sum_{j\in h} \sum_{n_j=1}^\infty 
	n_jz_{j(n_j)} \sum_{k \in h} \sum_{n_k=1}^\infty n_kz_{k(n_k)} 
	\frac{Z_{\vec{Q}-n_j\vec{q}_j-n_k\vec{q}_k}}{Z_{\vec{Q}}}\right]\nonumber \\
	& &+ 6\left[\sum_{j\in h} \sum_{n_j=1}^\infty 
	n_jz_{j(n_j)} \sum_{k \in h} \sum_{n_k=1}^\infty z_{k(n_k)}\sum_{l \in h} \sum_{n_l=1}^\infty z_{l(n_l)} 
	\frac{Z_{\vec{Q}-n_j\vec{q}_j-n_k\vec{q}_k-n_l\vec{q}_l}}{Z_{\vec{Q}}}\right]\nonumber\\
	& &+ \left[\sum_{j\in h} \sum_{n_j=1}^\infty 
	z_{j(n_j)} \sum_{k \in h} \sum_{n_k=1}^\infty z_{k(n_k)}\sum_{l \in h} \sum_{n_l=1}^\infty z_{l(n_l)}\sum_{m \in h} \sum_{n_m=1}^\infty z_{m(n_m)}\frac{Z_{\vec{Q}-n_j\vec{q}_j-n_k\vec{q}_k-n_l\vec{q}_l-n_m\vec{q}_m}}{Z_{\vec{Q}}}\right].
\end{eqnarray}
\end{widetext}

\section{Asymptotic fluctuations in the canonical ensemble}
\label{SecAsymptotic}

We describe fluctuations by various order moments of a multiplicity distribution and the products of the moments:
\begin{eqnarray}
	\frac{\sigma^2}{M}=\frac{\langle(\Delta N)^2\rangle}{\langle N \rangle},\hskip 0.5cm
	S\sigma=\frac{\langle(\Delta N)^3\rangle}{\langle(\Delta N)^2\rangle}, \nonumber\\{\rm and} \hskip 0.3cm 
	\kappa\sigma^2=\frac{\langle(\Delta N)^4\rangle-3\langle(\Delta N)^2\rangle}{\langle(\Delta N)^2\rangle},
\end{eqnarray}
where $N$ is the multiplicity of any hadron species and $\Delta N = N - \langle N \rangle$.

To get an idea about the extra-Poisson fluctuations in the CE, a simple illustrative example was introduced in Ref. \cite{ConservationBecattiniPRC2005}, by neglecting baryon number and strangeness, i.e., 
considering only electric charge. Also neglecting possible hadrons with two or more units of electric charge and within the classical Boltzmann approximation, the following asymptotic expression for the variance of a hadron species $j$ can be obtained \cite{ConservationBecattiniPRC2005}:  
\begin{equation}\label{SigmaM}
	\langle(\Delta N_j)^2\rangle  =  \langle N_j \rangle\left[1 - \frac{\langle N_j \rangle_{\rm GC}}
	{\langle h^+ \rangle_{\rm GC} + \langle h^- \rangle_{\rm GC}}\right] + \mathcal{O}(V^{-1}),
\end{equation}
where $\langle N_j \rangle_{\rm GC}$, $\langle h^+ \rangle_{\rm GC}$ and $\langle h^- \rangle_{\rm GC}$ are the mean multiplicities of hadron species $j$,  
positive and negative hadrons respectively in the GCE.
Eq.~(\ref{SigmaM}) show that, in the thermodynamic limit $V\rightarrow\infty$, the deviation from the Poisson statistics, $\langle(\Delta N_j)^2\rangle  =  \langle N_j \rangle$, is 
proportional to the relative content of multiplicity of the species $j$
with respect to all species carrying a non-vanishing value of the same charge.  
This is quite consistent with our physical 
intuitions that, in a system with many species, the single and rarely 
produced ones has small $\frac{\langle N_j \rangle_{\rm GC}}
{\langle h^+ \rangle_{\rm GC} + \langle h^- \rangle_{\rm GC}}$, and its distribution is little affected by global conservation laws. In the other extreme, the scaled variance of the most inclusive sets are strongly 
affected by the conservation laws and could deviate largely from Poisson statistics.

Following the same derivation as Ref.~\cite{ConservationBecattiniPRC2005}, within this simple model and the Boltzmann approximation, we obtain for the third and forth order cumulants:
\begin{equation}\label{SSigma}
	\langle(\Delta N_j)^3\rangle   =  \langle N_j \rangle\left[1 - 3 \frac{\langle N_j \rangle_{\rm GC}}
	{\langle h^+ \rangle_{\rm GC} + \langle h^- \rangle_{\rm GC}}\right] + \mathcal{O}(V^{-1})
\end{equation}
and
\begin{eqnarray}\label{KappaSigma}
	&&\langle(\Delta N_j)^4\rangle - 3\langle(\Delta N_j)^2\rangle \nonumber\\
	&&=  \langle N_j \rangle\left[1 - 7 \frac{\langle N_j \rangle_{\rm GC}}
	{\langle h^+ \rangle_{\rm GC} + \langle h^- \rangle_{\rm GC}}\right] + \mathcal{O}(V^{-1}).
\end{eqnarray}
Eqs. (\ref{SSigma}) and (\ref{KappaSigma}) suggest that, for the higher order cumulants, the deviation from the Poisson statistics is 
proportional to the same relative multiplicity weight as the second order, but with a larger coefficient, which indicates a larger canonical suppression of the higher order fluctuation. 
For the cumulants of the set of all positive and negative hadrons similar relations ban be obtained \cite{ConservationBecattiniPRC2005}, only modifying $N_j$ in Eqs. (\ref{SigmaM})-(\ref{KappaSigma}) to all positive and negative multiplicities respectively.

\begin{figure*}
	\begin{center}
		\hspace{-0.3cm}\includegraphics[width=9.8cm]{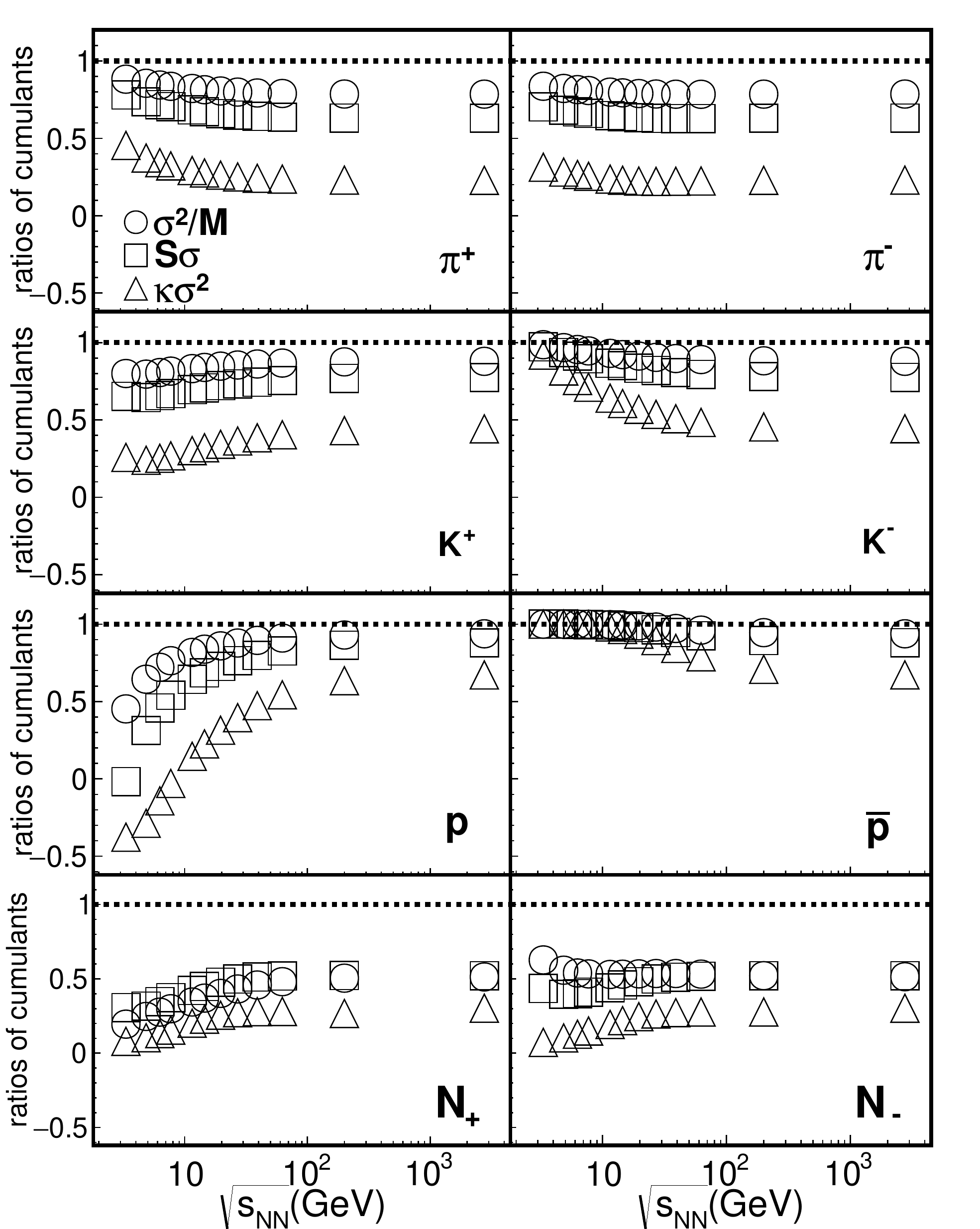}
		\hspace{-0.3cm}\includegraphics[width=5.6cm]{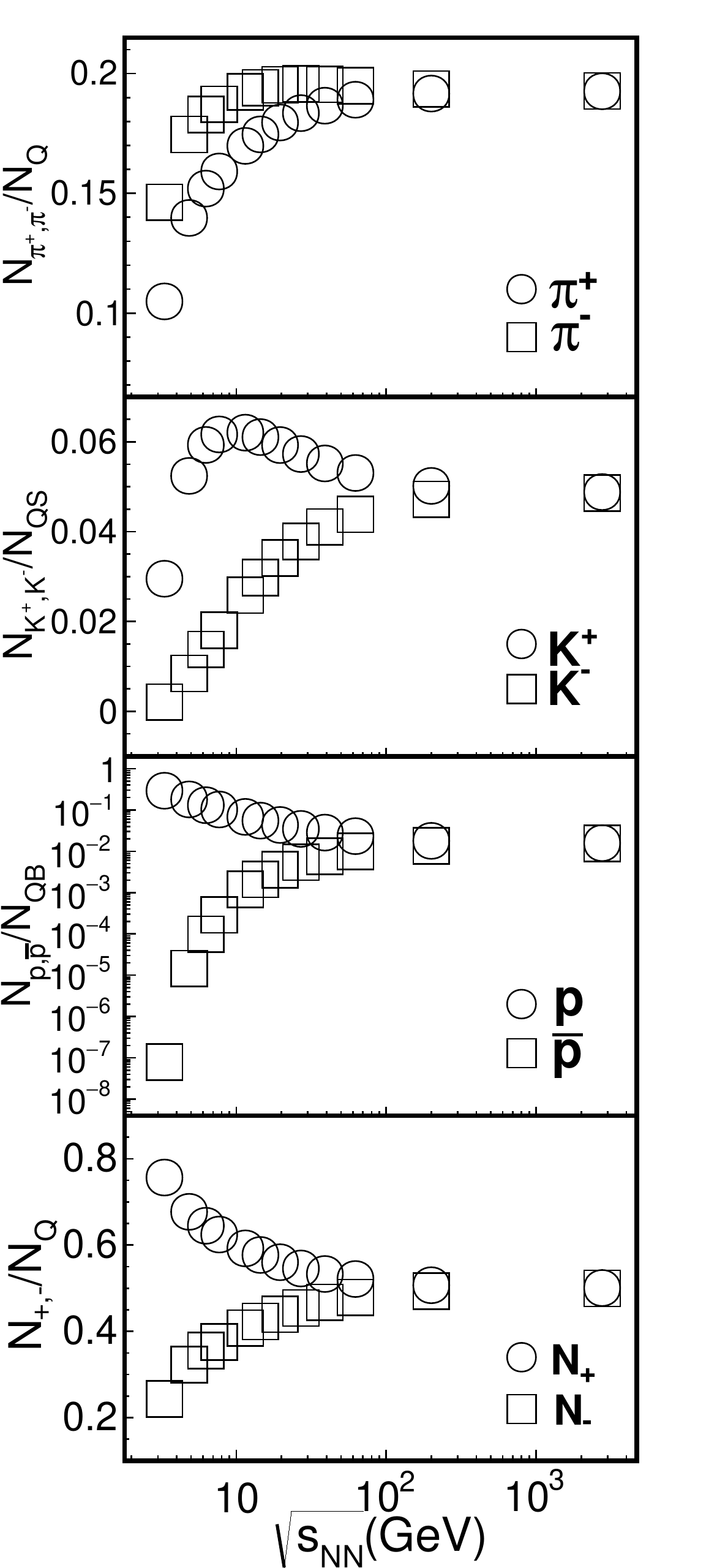}
		\caption{(left) The products of the moments, $\sigma^2/M$, $S\sigma$, and
			$\kappa\sigma^2$, of different sets of primary hadrons, $\pi^+$, $\pi^-$, $K^+$, $K^-$, $p$, $\bar{p}$, positive ($N_+$) and negative ($N_-$) charged particles, in the CE HRG
			model at SPS, RHIC and LHC energies. (right) Multiplicities of $\pi^+$, $\pi^-$, $K^+$, $K^-$, $p$, $\bar{p}$, positive and negative charged particles
			with respect to total multiplicity of all species carrying a non-vanishing value of the same charge, where $N_Q$ indicates multiplicity of particles with a nonzero electric charge, $N_{Q,S}$ a nonzero electric charge or strangeness, and $N_{Q,B}$ a nonzero electric charge or baryon number.} \label{FigParticleMoments}
	\end{center}
\end{figure*}

To test whether the expressions from the simplified model, Eqs. (\ref{SigmaM}) - (\ref{KappaSigma}), are valid in the full hadron gas, moment products, $\sigma^2/M$, $S\sigma$, and
$\kappa\sigma^2$, of different sets of primary hadrons, $\pi^+$, $\pi^-$, $K^+$, $K^-$, $p$, $\bar{p}$, positive ($N_+$) and negative ($N_-$) charged particles, are calculated in the CE HRG
model at SPS, RHIC and LHC energies, and the results are presented in Fig.~\ref{FigParticleMoments} (left). 
The results are obtained with an extended version of the thermal model program THERMUS, as explained in detail in the following Sec. \ref{SecFreezeout}. The dashed lines indicate the Poisson expectation. The right part of Fig.~\ref{FigParticleMoments} presents multiplicities of $\pi^+$, $\pi^-$, $K^+$, $K^-$, $p$, $\bar{p}$, positive and negative charged particles
with respect to total multiplicity of all species carrying a non-vanishing value of the same charge, where $N_Q$ indicates multiplicity of particles with a nonzero electric charge, $N_{Q,S}$ indicates multiplicity of particles with a nonzero electric charge or strangeness, and $N_{Q,B}$ a nonzero electric charge or baryon number.

From Fig. \ref{FigParticleMoments} we see that, qualitatively, the asymptotic relations Eqs. (\ref{SigmaM}) - (\ref{KappaSigma}) are approximately satisfied in the full hadron gas. 
Generally, higher order fluctuations in the CE have larger deviation (suppression) from the Poisson expectation, and the deviation is larger where the relative density of the species under study is larger. At low energies, the moments products of proton are largely suppressed, and $K^+$ suppresses more than $K^-$. As the collision energy increases, the particle and anti-particle densities approach each other and their moments products become similar. The second order fluctuation results obtained here are in good agreement with the previously reported values \cite{ConservationGorensteinPRC2005, ConservationBecattiniPRC2005}.
These asymptotic relations hold better for single hadron species pion, kaon and proton. For all positive and negative charged hadrons, the third and fourth order cumulants do not seem to suppress as much as those predicted by Eqs. (\ref{SSigma}) and (\ref{KappaSigma}).   
Nevertheless, these asymptotic relations can be taken as ``rules of thumb" to make estimates of the products of the moments to a first approximation \cite{ConservationBecattiniPRC2005}.

\section{Freeze-out conditions in heavy-ion collisions}
\label{SecFreezeout}
There is a phenomenological relation between the collision energy
and the corresponding thermal parameters, which defines the so
called chemical freeze-out line in the temperature and baryon
chemical potential plane \cite{Cleymans2006prc}. With increasing colliding
energy, the system temperature increases. This is accompanied by a
drop in baryon chemical potential, which can be parameterized by the
following function \cite{Cleymans2006prc}:
\begin{eqnarray}
	\mu_B\left(\sqrt{s_{NN}}\right)={1.308 \,{\rm
			GeV}}\big/\left(1+0.273 \,{\rm GeV^{-1}}\sqrt{s_{NN}}\right) ,\nonumber\\
\end{eqnarray}
where $\sqrt{s_{NN}}$ is the c. m. energy in units of GeV. The
chemical freeze-out temperature is determined by $\langle
E\rangle/\langle N\rangle \approx 1.08$ GeV \cite{Cleymans1998prl}, for energy per hadron.
Other model parameters, the chemical potentials related to
strangeness and isospin, are constrained by strangeness conservation
and by total charge over total baryon $Q/B=0.4$, respectively. The
strangeness saturation factor $\gamma_s$ has been set to 1. 

For the CE calculation, the needed intensive input parameters are the temperature $T$ and the charge densities.
The temperature and baryon density $\rho_B$ are determined from the GCE results at the same energy, and the values of $T$ and $\rho_B$ at different collision energies are listed in Table \ref{TableRhoB}.
The strangeness density $\rho_S$ is set to zero, and the electric charge density is set to $\rho_Q=0.4\rho_B$, 
corresponding to the ratio $Z/A$ of Pb + Pb and Au + Au collisions. 

The simulation can be carried out only with a finite volume. The computing time of the integral of Eq. (\ref{projection}) increases largely as the baryon number of the system increases \cite{Thermus}.  We choose volume $V = 1000~{\rm fm^3}$ for c. m. energy $\sqrt{s_{NN}}$ below 100 GeV and $V = 3000~{\rm fm^3}$ for c. m. energy above 100 GeV. Because of the large volume, which is used to ensure effective reaching of the thermodynamic limit \cite{ConservationBecattiniPRC2005,CleymansCanonical1997prc},  the efficiency of these runs is very low.

An extended version of the thermal model program THERMUS \cite{Thermus} is used in this
analysis. Quantum statistics and the finite width of resonances have
been taken into account. The
standard THERMUS particle table includes all mesons (up to the
$K_4^*(2045)$) and baryons (up to the $\Omega^-$) listed in the July 2002 Particle Physics Booklet \cite{PDG}, and their respective
decay channels. Weak decays are not included in the calculations.

\begin{table}[h!]
	\begin{center}
		\begin{tabular}{ccccccc}\hline
			\hline $\sqrt{s_{NN}}(\rm{GeV})$ & 3.32 & 4.85 & 6.27 & 7.7 & 11.5 & 14.5  \\ \hline
			$T$(${\rm MeV}$) & 94.7 & 117.9 & 129.9 & 137.8 & 149.0 & 153.4  \\ \hline
			$\rho_B$(${\rm fm}^{-3}$) & 0.093 & 0.112 & 0.118 & 0.118 & 0.107 & 0.097  \\ \hline
			\hline $\sqrt{s_{NN}}(\rm{GeV})$  & 19.6 & 27 & 39 & 62.4 & 200 & 2760\\ \hline
			$T$(${\rm MeV}$) & 157.4 & 160.1 & 161.9 & 163.0 & 163.7 & 163.8  \\ \hline
			$\rho_B$(${\rm fm}^{-3}$)& 0.081 & 0.065 & 0.048 & 0.032 & 0.010 & 0.0008\\
			\hline\hline
		\end{tabular}
		\caption{Temperature and baryon density at chemical freeze-out for central Au + Au (Pb + Pb)  collisions at SPS, RHIC and LHC energies.} \label{TableRhoB}
	\end{center}
\end{table}

\section{Effect of resonance decays}
\label{SecResonance}

In the HRG model, after thermal ``production", resonances and
heavier particles are allowed to decay, therefore contributing to
the final yields of lighter mesons and baryons. The ensemble
averaged final particle yields, after resonance decays, equal to
\cite{JeonRatioFlu1999, Begun06prc}
\begin{eqnarray}
	\langle N_i\rangle & = & \langle N_i^* \rangle + \sum_{R} \langle
	N_R \rangle \langle n_i \rangle_{R}, \label{resoyield}
\end{eqnarray}
where $N_i^*$ and $N_R$ denote the primordial yields of particles of
species $i$ and resonances $R$, the summation $\sum_{R}$ runs over
all types of resonances, and $\langle n_i \rangle_{R} \equiv \sum_r
b_r^R n_{i,r}^R$ is the average over resonance decay channels. The
parameter $b_r^R$ is the branching ratio of the $r$-th branch of
resonance $R$ decay, and $n_{i,r}^R$ is the number of particles of
species $i$ produced in the decay of resonance $R$ via the decay
mode $r$.

Resonance decay has a probabilistic character. This itself causes
the particle number fluctuations in the final state. The two
particle correlations after resonance decays can be calculated as
\cite{JeonRatioFlu1999, Begun06prc}
\begin{eqnarray}
	&&\langle \Delta N_i \Delta N_j \rangle = \langle \Delta N_i^*
	\Delta N_j^* \rangle +\sum_{R}\langle N_R \rangle \langle \Delta n_i
	\Delta n_j \rangle_{R} \nonumber\\&& + \sum_{R}\langle \Delta N_i^*\Delta N_R
	\rangle \langle n_j \rangle_{R} + \sum_{R}\langle
	\Delta N_j^*\Delta N_R \rangle \langle n_i \rangle_{R} \nonumber\\&& +
	\sum_{R,R'}\langle \Delta N_R\Delta N_{R'} \rangle \langle n_i
	\rangle_{R} \langle n_j \rangle_{R'}, \label{resofluct_CE}
\end{eqnarray}
where $\langle \Delta n_i \Delta n_j \rangle_{R} \equiv \sum_r b_r^R
n_{i,r}^R n_{j,r}^R - \langle n_i \rangle_R \langle n_j \rangle_R .$
In the CE, because of the presence of exact charge conservation laws, all primary particles and resonances correlate with one another, while, in the GCE, only particles of the same species do correlate, and Eq. (\ref{resofluct_CE}) reduces to  
\begin{eqnarray}
	&&\langle \Delta N_i \Delta N_j \rangle = \langle \Delta N_i^* \Delta
	N_j^* \rangle \nonumber\\&& +\sum_{R}\left[\langle (\Delta N_R)^2 \rangle \langle
	n_i \rangle_{R} \langle n_j \rangle_{R} + \langle N_R \rangle
	\langle \Delta n_i \Delta n_j \rangle_{R}\right]. \label{resofluct_GEC}
\end{eqnarray}

The three and four particle correlations after resonance decays can
be obtained similarly from the following generating function
\cite{Begun06prc}:
\begin{equation}
	G\equiv\prod_{R}\left(\sum_{r}b_r^R\prod_{i}\lambda_i^{n_{i.r}^R}\right)^{N_R},
\end{equation}
where $\lambda_i$ are auxiliary parameters that are set to one in
the final formula. The full results have been presented in Appendix A of Ref. \cite{Jinghua2013PLB}, and to save space not repeated here.
To calculate the resonance decay contributions to kurtosis in the CE include four nested loops over all particles and resonances, which is again fairly computing time consuming.

In recent studies \cite{Nahrgang2015EPJC,Mohanty2016PRC}, the resonance decay contributions have been divided into an average and a probabilistic part. For example, in Eq. (\ref{resofluct_GEC}), the second term on the right hand side is called the average part, and the third term called the probabilistic part. The combination of these two parts is called the full contribution. The resonance decay contribution in the present analysis is not separated. 
Wherever it is  mentioned, it always refers to the full contribution.

\section{Moments of net-proton, net-kaon and net-charge multiplicity distributions on the chemical freeze-out curve}
\label{SetNetparticle}

The net-particle quantities, net-proton, net-kaon, and net-charge, are formed event-by-event as, $N_{p-\bar{p}}=N_p-N_{\bar{p}}$,  $N_{K^+-K^-}=N_{K^+}-N_{K^-}$, and $N_{\rm net-charge}=N_{+}-N_{-}$. The cumulants of net-proton
multiplicity distributions can be obtained from the cumulants of
proton and anti-proton multiplicity distributions and their
correlations:
\begin{eqnarray}
M&=&\langle N_{p-\bar{p}}\rangle=\langle N_p \rangle - \langle
N_{\bar{p}} \rangle , \label{npmean}\\
\sigma^2&=&\langle (\Delta N_{p-\bar{p}})^2 \rangle\nonumber\\
&=&\langle (\Delta
N_p)^2 \rangle + \langle (\Delta N_{\bar{p}})^2 \rangle-2\langle\Delta N_p\Delta N_{\bar{p}}\rangle, \label{npvariance}
\end{eqnarray}
\begin{eqnarray}
\langle (\Delta N_{p-\bar{p}})^3\rangle &=& \langle (\Delta N_p)^3
\rangle - 3\langle (\Delta N_p)^2(\Delta N_{\bar{p}}) \rangle \nonumber\\
&&+ 3\langle (\Delta N_p)(\Delta N_{\bar{p}})^2 \rangle
- \langle (\Delta N_{\bar{p}})^3 \rangle , \label{npc3} 
\end{eqnarray}
%\begin{widetext}
\begin{eqnarray}
&&\langle(\Delta N_{p-\bar{p}})^4\rangle-3\langle(\Delta
	N_{p-\bar{p}})^2\rangle^2\nonumber\\&&=\left[\langle(\Delta
N_p)^4\rangle-3\langle(\Delta N_p)^2\rangle^2\right]
+\left[\langle(\Delta N_{\bar{p}})^4\rangle-3\langle (\Delta
N_{\bar{p}})^2\rangle^2\right]\nonumber\\
&&-4\left[\langle(\Delta N_p)^3 (\Delta N_{\bar{p}})\rangle
-3\langle(\Delta N_p)^2\rangle\langle\Delta N_p\Delta
N_{\bar{p}}\rangle\right]\nonumber\\
&&+6\left[\langle(\Delta N_p)^2(\Delta N_{\bar{p}})^2\rangle
-2\langle\Delta N_p\Delta N_{\bar{p}}\rangle^2-\langle(\Delta
N_p)^2\rangle\langle(\Delta N_{\bar{p}})^2\rangle\right]\nonumber\\
&&-4\left[\langle(\Delta N_p)(\Delta N_{\bar{p}})^3\rangle
-3\langle(\Delta N_{\bar{p}})^2\rangle\langle\Delta N_p\Delta
N_{\bar{p}}\rangle\right].\label{npc4}
\end{eqnarray}
%\end{widetext}

The cumulants of proton, anti-proton and their correlations in the CE can be calculated according to Eqs. (\ref{CEmean}) - (\ref{CEcumulant4}). Formulas for cumulants in the GCE can be found in previous publications \cite{Jinghua2013PLB}.
The cumulants of net-kaon and net-charge are obtained similarly from the cumulants of $K^+$ and $K^-$ and the cumulants of positive and negative charged particles, respectively. 

\begin{figure*}
	\begin{center}
		\hspace{-0.2cm}\includegraphics[width=5.5cm]{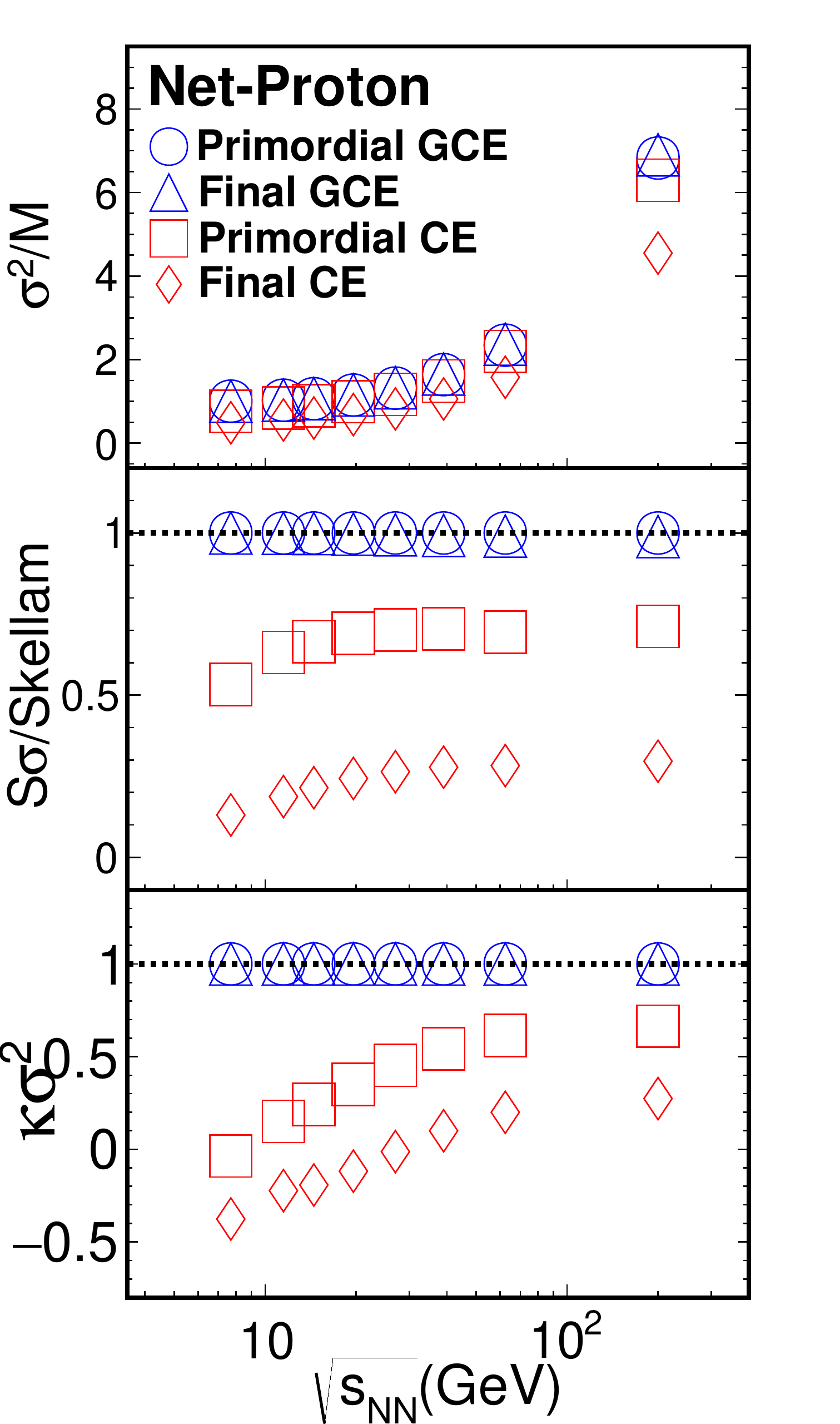}
		\hspace{-0.4cm}\includegraphics[width=5.5cm]{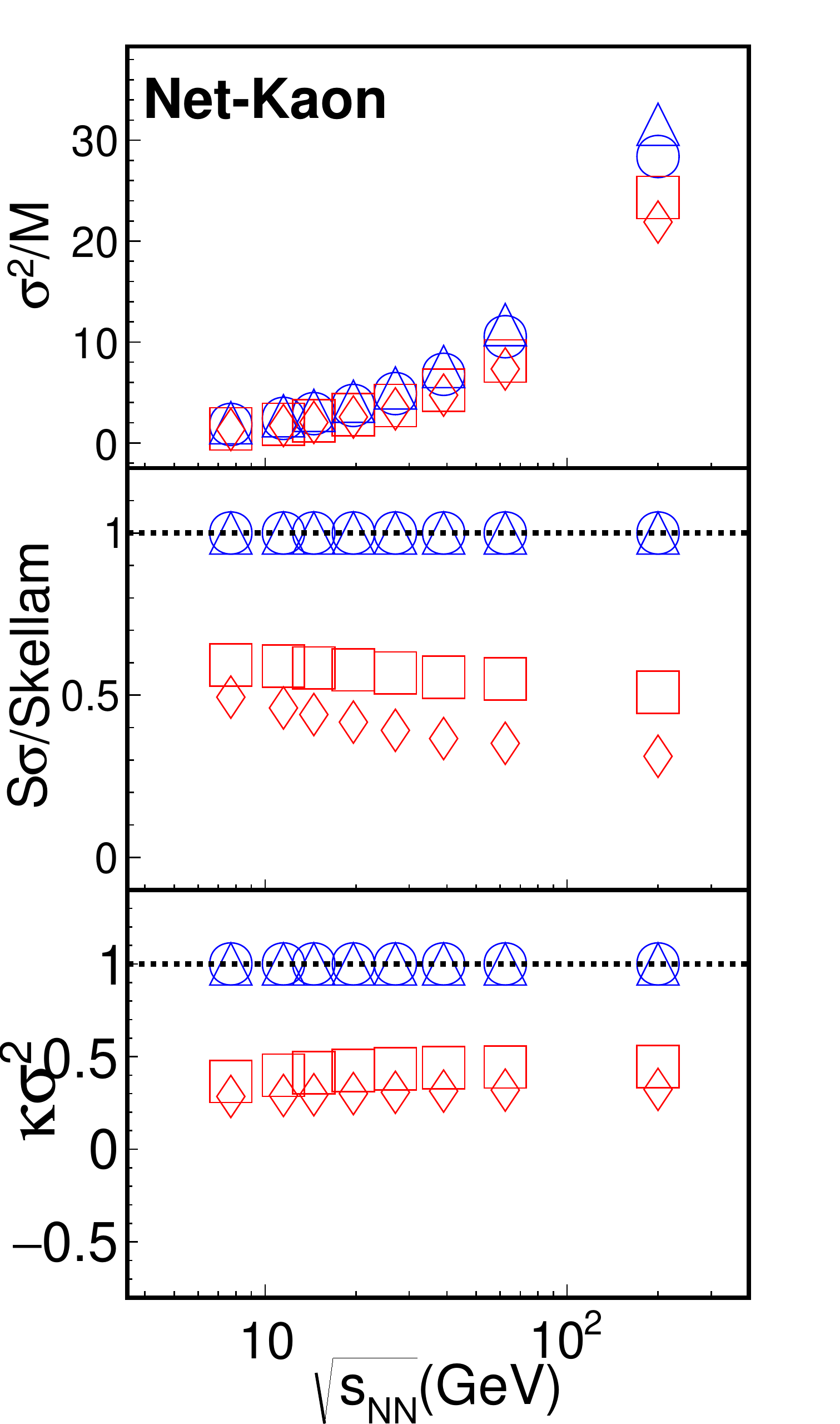}
		\hspace{-0.4cm}\includegraphics[width=5.5cm]{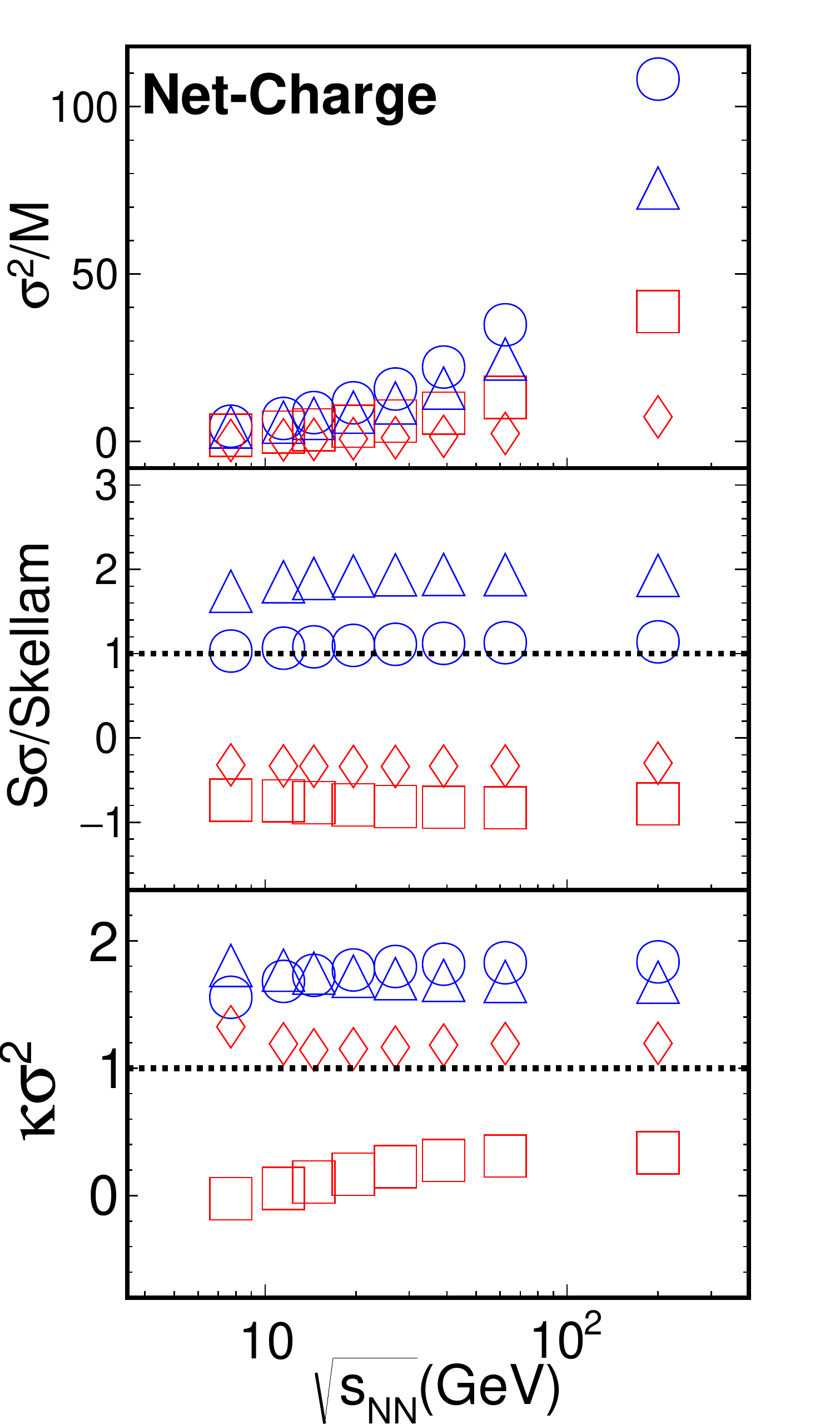}
		\caption{Energy dependence of the products of the moments, $\sigma^2/M$, $S\sigma/\rm{Skellam}$,  and
			$\kappa\sigma^2$, for net-proton (left figure), net-kaon (middle figure), and net-charge (right figure) distributions from the HRG model. } \label{FigNetParticle}
	\end{center}
\end{figure*}

Energy dependence of the products of the moments, $\sigma^2/M$, $S\sigma/\rm{Skellam}$, and
$\kappa\sigma^2$, of net-proton, net-kaon, and net-charge distributions calculated in the HRG
model at RHIC energies are
presented in Fig. \ref{FigNetParticle}. The model results are presented for primordial and final state particles in both GCE and CE at $\sqrt{s_{NN}}=7.7, 11.5, 14.5, 19.6, 27,
39, 62.4,$ and 200 GeV. The relevant primordial and final state values of moments products for net-proton, net-kaon, and net-charge for various colliding energies are summarized in Tables \ref{TablePrimProton} - \ref{TableFinalCharge}, respectively. 

\begin{table}[h!]
	\begin{center}
		\begin{tabular}{||c||c|c||c|c||c|c||}\hline
			& \multicolumn{2}{c||}{ $\sigma^2/M$} & \multicolumn{2}{c||}{
				$S\sigma/\rm{Skellam}$} & \multicolumn{2}{c||} { $\kappa\sigma^2$} \\ [0.5ex]
			\hline $\sqrt{s_{NN}} (\rm{GeV})$ & GCE & CE & GCE & CE&
			GCE & CE \\ [0.5ex] \hline\hline
			$7.7$ & 0.995 & 0.764 & 0.984 & 0.533 & 0.952 & -0.036 \\
			$11.5$ & 1.022 & 0.839 & 0.989 & 0.632 & 0.969 & 0.151 \\
			$14.5$ & 1.059 & 0.890 & 0.991 & 0.665 & 0.975 & 0.242 \\
			$19.6$ & 1.147 & 0.991 & 0.992 & 0.691 & 0.982 & 0.350 \\
			$27$ & 1.314 & 1.166 & 0.993 & 0.701 & 0.986 & 0.453 \\
			$39$ & 1.638 & 1.486 & 0.993 & 0.704 & 0.990 & 0.543 \\
			$62.4$ & 2.344 & 2.195 & 0.994 & 0.694 & 0.992 & 0.614 \\
			$200$ & 6.828  & 6.296 & 0.994 & 0.712 & 0.993 & 0.664 \\
			\hline
		\end{tabular}
		\caption{Primordial net-proton $\sigma^2/M$, $S\sigma/\rm{Skellam}$, and
			$\kappa\sigma^2$ in the GCE and CE HRG model for central Au + Au (Pb + Pb) collisions.} \label{TablePrimProton}
	\end{center}
\end{table}

\begin{table}[h!]
	\begin{center}
		\begin{tabular}{||c||c|c||c|c||c|c||}\hline
			& \multicolumn{2}{c||}{ $\sigma^2/M$} & \multicolumn{2}{c||}{
				$S\sigma/\rm{Skellam}$} & \multicolumn{2}{c||} { $\kappa\sigma^2$} \\ [0.5ex]
			\hline $\sqrt{s_{NN}} (\rm{GeV})$ & GCE & CE & GCE & CE&
			GCE & CE \\ [0.5ex] \hline\hline
			$7.7$ & 1.000 & 0.486 & 0.992 & 0.131 & 0.976 & -0.376 \\
			$11.5$ & 1.026 & 0.553 & 0.994 & 0.187 & 0.986 & -0.223 \\
			$14.5$ & 1.063 & 0.600 & 0.994 & 0.215 & 0.990 & -0.193\\
			$19.6$ & 1.153 & 0.682 & 0.993 & 0.243 & 0.993 & -0.118 \\
			$27$ & 1.323 & 0.817 & 0.992 & 0.264 & 0.995 & -0.013 \\
			$39$ & 1.651 & 1.056 & 0.990 & 0.278 & 0.996 & 0.099 \\
			$62.4$ & 2.369 & 1.576 & 0.988 & 0.283 & 0.997 & 0.199 \\
			$200$ & 6.905  & 4.550 & 0.986 & 0.296 & 0.997 & 0.273 \\
			\hline
		\end{tabular}
		\caption{Final net-proton $\sigma^2/M$, $S\sigma/\rm{Skellam}$, and
			$\kappa\sigma^2$ in the GCE and CE HRG model for central Au + Au (Pb + Pb) collisions.} \label{TableFinalProton}
	\end{center}
\end{table}

\begin{table}[h!]
	\begin{center}
		\begin{tabular}{||c||c|c||c|c||c|c||}\hline
			& \multicolumn{2}{c||}{ $\sigma^2/M$} & \multicolumn{2}{c||}{
				$S\sigma/\rm{Skellam}$} & \multicolumn{2}{c||} { $\kappa\sigma^2$} \\ [0.5ex]
			\hline $\sqrt{s_{NN}} (\rm{GeV})$ & GCE & CE & GCE & CE&
			GCE & CE \\ [0.5ex] \hline\hline
			$7.7$ & 1.844 & 1.389 & 1.055 & 0.593 & 1.083 & 0.366 \\
			$11.5$ & 2.444 & 1.840 & 1.069 & 0.589 & 1.090 & 0.340 \\
			$14.5$ & 2.920 & 2.200 & 1.075 & 0.584 & 1.092 & 0.413 \\
			$19.6$ & 3.730 & 2.807 & 1.080 & 0.578 & 1.092 & 0.425 \\
			$27$ & 4.909 & 3.713 & 1.084 & 0.568 & 1.092 & 0.434 \\
			$39$ & 6.822 & 5.227 & 1.086 & 0.555 & 1.092 & 0.440 \\
			$62.4$ & 10.557 & 8.114 & 1.087 & 0.550 & 1.091 & 0.444 \\
			$200$ & 28.558  & 24.330 & 1.088 & 0.508 & 1.091 & 0.447 \\
			\hline
		\end{tabular}
		\caption{Primordial net-kaon $\sigma^2/M$, $S\sigma/\rm{Skellam}$, and
			$\kappa\sigma^2$ in the GCE and CE HRG model for central Au + Au (Pb + Pb) collisions.} \label{TablePrimKaon}
	\end{center}
\end{table}

\begin{table}[h!]
	\begin{center}
		\begin{tabular}{||c||c|c||c|c||c|c||}\hline
			& \multicolumn{2}{c||}{ $\sigma^2/M$} & \multicolumn{2}{c||}{
				$S\sigma/\rm{Skellam}$} & \multicolumn{2}{c||} { $\kappa\sigma^2$} \\ [0.5ex]
			\hline $\sqrt{s_{NN}} (\rm{GeV})$ & GCE & CE & GCE & CE&
			GCE & CE \\ [0.5ex] \hline\hline
			$7.7$ & 2.012 & 1.328 & 1.039 & 0.494 & 1.056 & 0.285 \\
			$11.5$ & 2.702 & 1.719 & 1.045 & 0.461 & 1.055 & 0.290 \\
			$14.5$ & 3.241 & 2.034 & 1.047 & 0.440 & 1.054 & 0.294 \\
			$19.6$ & 4.153 & 2.567 & 1.048 & 0.417 & 1.053 & 0.300 \\
			$27$ & 5.474 & 3.372 & 1.049 & 0.392 & 1.051 & 0.307 \\
			$39$ & 7.616 & 4.734 & 1.049 & 0.366 & 1.051 & 0.314 \\
			$62.4$ & 11.792 & 7.330 & 1.050 & 0.352 & 1.050 & 0.319 \\
			$200$ & 31.794 & 21.891 & 1.050 & 0.311 & 1.050 & 0.323 \\
			\hline
		\end{tabular}
		\caption{Final net-kaon $\sigma^2/M$, $S\sigma/\rm{Skellam}$, and
			$\kappa\sigma^2$ in the GCE and CE HRG model for central Au + Au (Pb + Pb) collisions.} \label{TableFinalKaon}
	\end{center}
\end{table}

\begin{table}[h!]
	\begin{center}
		\begin{tabular}{||c||c|c||c|c||c|c||}\hline
			& \multicolumn{2}{c||}{ $\sigma^2/M$} & \multicolumn{2}{c||}{
				$S\sigma/\rm{Skellam}$} & \multicolumn{2}{c||} { $\kappa\sigma^2$} \\ [0.5ex]
			\hline $\sqrt{s_{NN}} (\rm{GeV})$ & GCE & CE & GCE & CE&
			GCE & CE \\ [0.5ex] \hline\hline
			$7.7$ & 4.526 & 1.836 & 1.031 & -0.736 & 1.556 & -0.024 \\
			$11.5$ & 6.834 & 2.763 & 1.065 & -0.746 & 1.681 & 0.057 \\
			$14.5$ & 8.597 & 3.431 & 1.081 & -0.782 & 1.729 & 0.106 \\
			$19.6$ & 11.514 & 4.531 & 1.098 & -0.794 & 1.771 & 0.168 \\
			$27$ & 15.641 & 6.045 & 1.111 & -0.813 & 1.798 & 0.226 \\
			$39$ & 22.206 & 8.413 & 1.122 & -0.820 & 1.816 & 0.276 \\
			$62.4$ & 34.825 & 13.096 & 1.130 & -0.828 & 1.827 & 0.311 \\
			$200$ & 108.24  & 38.775 & 1.140 & -0.829 & 1.834 & 0.336 \\
			\hline
		\end{tabular}
		\caption{Primordial net-charge $\sigma^2/M$, $S\sigma/\rm{Skellam}$, and
			$\kappa\sigma^2$ in the GCE and CE HRG model for central Au + Au (Pb + Pb) collisions.} \label{TablePrimCharge}
	\end{center}
\end{table}

\begin{table}[h!]
	\begin{center}
		\begin{tabular}{||c||c|c||c|c||c|c||}\hline
			& \multicolumn{2}{c||}{ $\sigma^2/M$} & \multicolumn{2}{c||}{
				$S\sigma/\rm{Skellam}$} & \multicolumn{2}{c||} { $\kappa\sigma^2$} \\ [0.5ex]
			\hline $\sqrt{s_{NN}} (\rm{GeV})$ & GCE & CE & GCE & CE&
			GCE & CE \\ [0.5ex] \hline\hline
			$7.7$ & 4.305 & 0.273 & 1.740 & -0.319 & 1.806 & 1.325 \\
			$11.5$ & 5.737 & 0.454 & 1.853 & -0.331 & 1.769 & 1.191 \\
			$14.5$ & 6.846 & 0.580 & 1.892 & -0.335 & 1.746 & 1.143 \\
			$19.6$ & 8.727 & 0.783 & 1.921 & -0.339 & 1.721 & 1.152 \\
			$27$ & 11.460 & 1.061 & 1.935 & -0.338 & 1.702 & 1.165 \\
			$39$ & 15.904 & 1.500 & 1.939 & -0.333 & 1.689 & 1.181 \\
			$62.4$ & 24.584 & 2.352 & 1.937 & -0.332 & 1.680 & 1.193 \\
			$200$ & 75.705  & 7.299 & 1.928 & -0.298 & 1.674 & 1.195 \\
			\hline
		\end{tabular}
		\caption{Final net-charge $\sigma^2/M$, $S\sigma/\rm{Skellam}$, and
			$\kappa\sigma^2$ in the GCE and CE HRG model for central Au + Au (Pb + Pb) collisions.} \label{TableFinalCharge}
	\end{center}
\end{table}

From Fig. \ref{FigNetParticle}, the GCE HRG results of $\sigma^2/M$ of net-proton, net-kaon, and net-charge distributions increase with collision energy.
As collision energy increases, the densities and variances of both particle and anti-particle increase, 
in the meanwhile their differences decrease. The variance of the net-particle is the sum of the particle and anti-particle variance, while the density of the net-particle is the difference of the particle and anti-particle density, as shown in Eqs. (\ref{npmean}) and (\ref{npvariance}), which leads to the increase of net-particle $\sigma^2/M$ with collision energy. 

Since there is no resonance decays into both proton and anti-proton, the resonance decay contribution to the net-proton fluctuations in the GCE is negligible. The effect of resonance decays remains small for the net-kaon fluctuations in the GCE, but it is important for the net-charge fluctuations. Resonance decays suppress the net-charge fluctuation, $\sigma^2/M$ decreasing especially at high RHIC energies. This is quite different from the total-charge fluctuation, where resonance decays increase its $\sigma^2/M$ \cite{Begun06prc,ConservationBecattiniPRC2005}. Resonance decays lead to large positive correlations between positive and negative charged particles, $\langle\Delta N_+ \Delta N_-\rangle$. This term, the last term of Eq. (\ref{npvariance}),  contributes positively to the total-charge variance, but negatively to the net-charge variance, which leads to the relative reduction of the net-charge variance when include resonance decays, and $\sigma^2/M$ decreasing. The effect of resonance decays increases with collision energy as expected.

The net-proton and net-kaon $\sigma^2/M$ from the CE HRG are mildly suppressed compared with the GCE results, while the net-charge $\sigma^2/M$ from the CE HRG are strongly suppressed. The strong suppression of the net-charge $\sigma^2/M$ in the CE reflects charge conservation. When resonance decays included, the final state net-charge $\sigma^2/M$ from the CE HRG is approaching zero because when all the charged particles are included the net charge ($Q$) is exactly conserved in the CE, and there is no fluctuation. 

The GCE HRG results of $S\sigma/\rm{Skellam}$ and $\kappa\sigma^2$ of primordial net-proton and net-kaon are approximately one except for quantum statistical corrections. The results of net-proton are slightly suppressed by Fermi statistics, and those of net-kaon are slightly enhanced by Bose statistics. The quantum statistical corrections for net-charge is larger, because it has contributions from all stable charged particles, especially those of pion. To be consistent with experimental measurements, net-charge distributions in Fig. \ref{FigNetParticle} and the following Fig. \ref{FigDeltaEta} take into account only stable charged particles, while positive and negative charged particles in Fig. \ref{FigParticleMoments} include all primary charged hadrons. 
Stable charged particles here include $\pi^+$, $\pi^-$, $K^+$, $K^-$ as well as $p$, $\Sigma^+$, $\Sigma^-$, $\Xi^-$, $\Omega^-$ and their respective anti-baryons. 

The CE HRG results of $S\sigma/\rm{Skellam}$ and $\kappa\sigma^2$ of net-proton and net-kaon are suppressed compared with the GCE results. The suppression of $\kappa\sigma^2$ is larger than $S\sigma/\rm{Skellam}$, consistent with our previous observations for single particle species, higher order cumulants more suppressed in the CE. For net-proton, the canonical suppression is larger at low energies, due to the large proton density and large suppression of proton cumulants at low energies.
The canonical suppression of net-kaon $S\sigma/\rm{Skellam}$ and $\kappa\sigma^2$  depend weakly on the collision energy. The net-kaon $S\sigma/\rm{Skellam}$ is slightly more suppressed at high RHIC energies, because the $K^+$ and $K^-$ skewness become close to each other at high energies, and the net-kaon skewness decreases. The net-kaon $\kappa\sigma^2$ is more suppressed at low RHIC energies, which is mainly due to the large suppression of $K^+$ kurtosis at low energies. 

The primordial and final net-charge $S\sigma/\rm{Skellam}$ in the CE become negative at all RHIC energies. The net-charge skewness gets positive contribution from the $N_+$ skewness and negative contribution from the $N_-$ skewness. In the CE, the skewness of $N_+$ becomes smaller than the skewness of $N_-$, because the density of $N_+$ is larger than the density of $N_-$ and the higher order cumulants of $N_+$ are more suppressed. 

Including resonance decays further suppresses net-proton and net-kaon $S\sigma/\rm{Skellam}$ and $\kappa\sigma^2$ in the CE, but increases net-charge  $S\sigma/\rm{Skellam}$ and $\kappa\sigma^2$ in the CE.  Generally, conservation laws  suppress particle production and correlation, while resonance decays increase particle production and correlation. The final results reflect a combination of these two effects. The resonance decay effect is smallest for net-proton,  largest for net-charge, and net-kaon stay in between of them.   

The STAR experiment measured results of net-proton $S\sigma/\rm{Skellam}$ and $\kappa\sigma^2$  \cite{LuoXFCPOD2014} with small $p_T$ acceptance, $0.4<p_T<0.8$ GeV/c,  are approximately 1, similar to the GCE HRG results. The STAR results of net-proton $S\sigma/\rm{Skellam}$ with larger $p_T$ acceptance, $0.4<p_T<2$ GeV/c,  are suppressed compared with the smaller $p_T$ acceptance results, and the suppression becomes larger for decreasing collision energy. This is quite similar to the CE HRG results with larger suppression 
at low energies due to the larger suppression of proton fluctuations in the CE at low energies. The STAR measured net-proton $\kappa\sigma^2$ with larger $p_T$ acceptance strongly increases at low energies, which is completely different from the strongly suppressed CE HRG results.  

\section{Pseudo-rapidity coverage dependence of net-charge fluctuation}
\label{SecRapCover}

\begin{figure}
	\begin{center}
		\includegraphics[width=6.2cm]{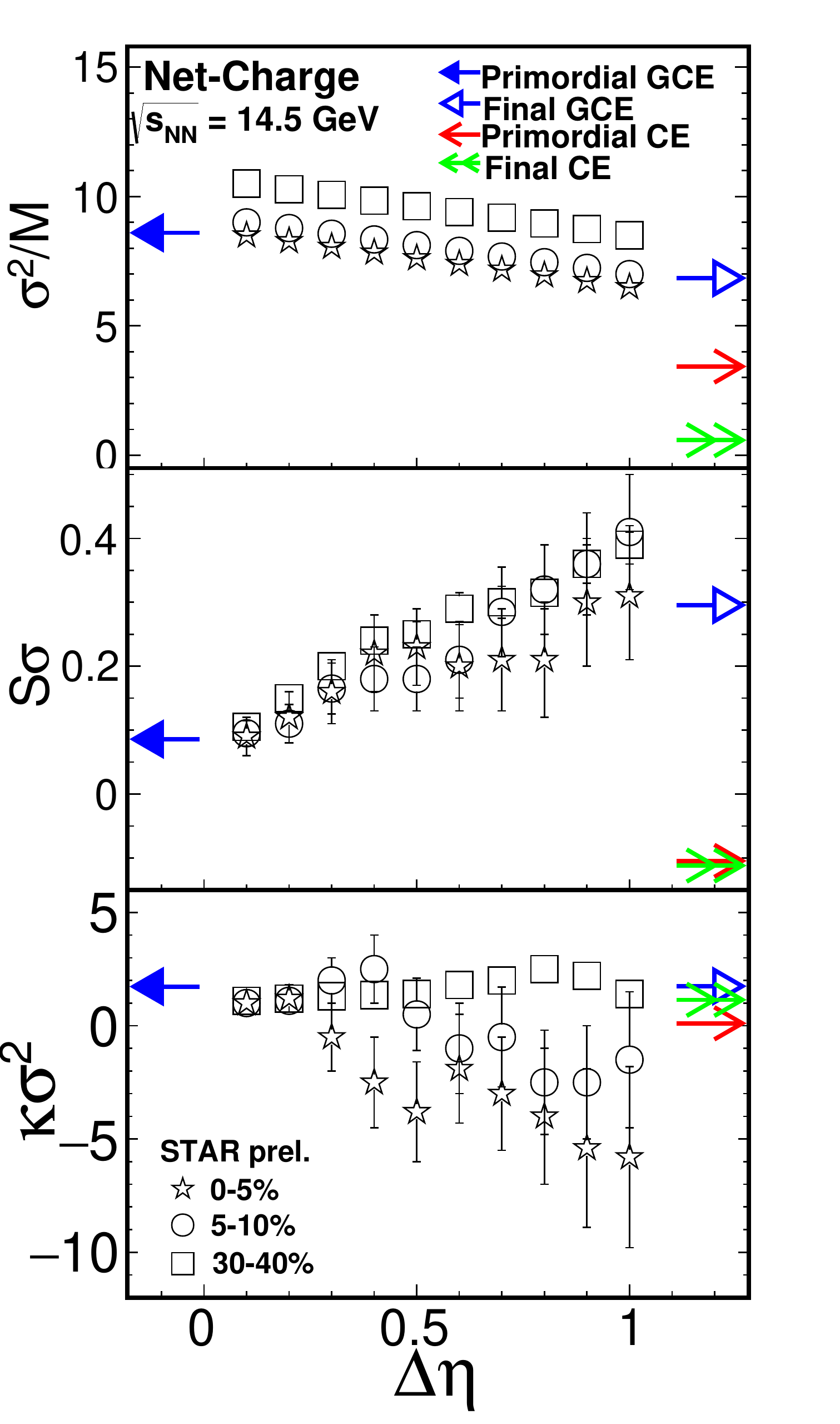}
		\caption{HRG model results for primordial and final state particles in both GCE and CE shown as different kinds of arrows compared with pseudo-rapidity coverage dependence of moments products, $\sigma^2/M$, $S\sigma$, and
			$\kappa\sigma^2$, for net-charge distributions obtained by STAR experiment \cite{JochenQM2015} at $\sqrt{s_{NN}}=14.5$ GeV. } \label{FigDeltaEta}
	\end{center}
\end{figure}

The $\eta$ coverage ($\Delta\eta$) dependence of net-charge fluctuation at $\sqrt{s_{NN}}=14.5$ GeV have been presented by the STAR Collaboration in Ref. \cite{JochenQM2015}.  Both effects of conservation laws and resonance decays become important as phase space coverage increases. To show how these two effects change the model calculations, 
the HRG results for net-charge distributions at $\sqrt{s_{NN}}=14.5$ GeV are indicated as different kinds of arrows in Fig. \ref{FigDeltaEta}. The model results for primordial particles in the GCE are shown as full blue arrows, and they are placed on the left sides of the figures to indicate that these model results are better for measurements with small acceptance . The primordial particle results in the GCE are supposed for measurements in a restricted phase space. The smaller the fraction of observed particles the smaller is the effect of global charge conservation, and the smaller is the correlation introduced by resonance decays. The model results for final state particles in the GCE and for primordial and final state particles in the CE are shown as open blue arrows, red arrows and double green arrows respectively. They are placed on the right sides of the figures to indicate that these model results are more suitable for measurements with large (full) acceptance. The STAR preliminary results are plotted in Fig. \ref{FigDeltaEta}.  
 
To compare the model results with experimental data, the limited detector acceptance should be taken into account. In the limit of a very small acceptance window,  one can assume uncorrelated particle detection, and the acceptance corrections may be modeled by a binomial distribution\cite{Gorenstein2007prc, Koch2013PRC}. For a particle with probability $p$ to be accepted, for small $p$, one obtains \cite{Begun2016prc}
\begin{align}\label{acceptance}
\omega &=  1+p(\omega_{4\pi}-1),\nonumber\\
S\sigma &=  1+2p(\omega_{4\pi}-1) + \mathcal{O}\left(p^{2}\right),\nonumber \\
\kappa\sigma^2 & =  1+6p(\omega_{4\pi}-1) + \mathcal{O}\left(p^{2}\right),
\end{align}
where $\omega=\sigma^2(N)/M$, and $\omega_{4\pi}$ is the scaled variance of the full acceptance. 
From Eq. (\ref{acceptance}), if the underlying distribution has $\omega_{4\pi}=1$, e.g., Poisson distribution, the measured moments products are always unity. The primordial particle production in the GCE with classical Boltzmann statistics belongs to this case. If the underlying $\omega_{4\pi}>1$, e.g., bosons with quantum statistics, the measured fluctuations are greater than unity. When global conservation laws are considered, fluctuations are suppressed, $\omega_{4\pi}<1$, the measured fluctuations are less than unity. The measured distribution approaches the Poisson one from above or below in the very small acceptance ($p \rightarrow 0$) limit. For large acceptance ($p \rightarrow1$) ,  independent production assumption no longer valid. In statistical models, correlations are caused by resonance decays, conservation laws and quantum statistics.  In recent studies, the effects of kinematic cuts are studied by modifying the lower and upper limits  of the integral of partition function \cite{Nahrgang2014PLB,KarschRedlich2016PRC}.  More studies are needed to improve the modeling of the limited experimental acceptance.   

It is not trying to make a detailed comparison between model and data here, but to show how resonance decays and conservation laws change the model results. 
At $\sqrt{s_{NN}}=14.5$ GeV, resonance decays in the GCE decrease net-charge $\sigma^2/M$ by about $20\%$, increase its $S\sigma$ by about $220\%$,  and leave $\kappa\sigma^2$ almost unchanged. As explained before, resonance decays lead to large positive correlations between positive and negative charge particles. These correlations contribute negatively to the net-charge $\sigma^2$, which reduces $\sigma^2/M$, and increases $S\sigma=\langle (\Delta N)^3\rangle/\sigma^2$. The large increase of $S\sigma$ is also caused by the increase of the third order cumulant, $\langle (\Delta N)^3\rangle$, when include resonance decays. The modifications of the forth order cumulant due to resonance decays are similar to those of the second order, and make $\kappa\sigma^2$ almost unchanged.    

For the smallest $\Delta\eta$ window, the STAR results are close to the primordial GCE results. This is the small acceptance limit, and the measured distribution shall approach the Poisson expectation. Due to quantum statistics, the GCE primordial net-charge $\kappa\sigma^2 \approx 1.7$. The STAR measured $\kappa\sigma^2$ is less than that and very close to one, which might be a reflection of the small acceptance effect, the measured fluctuations approaching unity from above. When $\Delta\eta$ increases, STAR measured $\sigma^2/M$ monotonically decreasing and $S\sigma$ monotonically increasing, which is quite consistent with the effects caused by resonance decays.  For $\sigma^2/M$, include conservation laws further decreases the model results, both resonance decays and conservation laws decrease the model results. For $S\sigma$, resonance decays and conservation laws change the model results in different direction. Resonance decays increase  $S\sigma$, while conservation laws decrease $S\sigma$. Notice, in Fig. \ref{FigNetParticle}, $S\sigma/\rm{Skellam}$ is presented, while, in Figs. \ref{FigParticleMoments} and \ref{FigDeltaEta}, it is simply $S\sigma$. The STAR measured $\kappa\sigma^2$ dependence on $\Delta\eta$ depends on centrality. For peripheral collisions, the STAR measured $\kappa\sigma^2$ slightly increases with $\Delta\eta$. This increasing might be a reflection of the quantum statistics or resonance decay effects, both giving similar results. For the most central collisions, the STAR measured $\kappa\sigma^2$ decreases with increasing $\Delta\eta$ with large error bars. Including conservation laws slightly decreases the net-charge $\kappa\sigma^2$, but less than the decreasing shown in data.

\section{Conclusion}
\label{SecConclusion}

The higher order moments of multiplicity distributions have been calculated in the  CE in a general multispecies hadron resonance gas. Exact conservation of three charges, baryon number, electric charge, and strangeness, has been enforced in the large volume limit. 
Moments products up to the forth order of various particles are calculated at SPS, RHIC and LHC energies. The asymptotic fluctuations within a simplified model in the CE are discussed where simple analytical expressions for moments of multiplicity distributions can be obtained. Both analytical results from the simplified model and the full HRG simulation results indicate that particle fluctuations in the CE are suppressed, higher order fluctuations generally have larger deviation (suppression) from the Poisson expectation, and the deviation is larger where the relative density of the species under study is larger.

The products of the moments of net-proton, net-kaon, and net-charge distributions in Au + Au collisions at RHIC energies are calculated within the HRG
model in the CE. The GCE results are calculated and presented as a comparison with the CE results.
The moments and their products are evaluated
on the phenomenologically determined freeze-out curve in the
temperature and baryon chemical potential plane. Quantum statistics and
resonance decay contributions have been taken into account in the model calculations. 

The HRG results of $\sigma^2/M$ of net-proton, net-kaon, and net-charge distributions increase with collision energy. Conservation laws suppress $\sigma^2/M$, and the suppression is largest for net-charge fluctuations. The primordial and final GCE results of $S\sigma/\rm{Skellam}$ and $\kappa\sigma^2$ for net-proton and net-kaon are approximately one, while those of net-charge are more sensitive to quantum statistics and resonance decays. The CE results of $S\sigma/\rm{Skellam}$ and $\kappa\sigma^2$ for net-proton, net-kaon and net-charge are suppressed compared with their GCE results. For net-proton the suppression is larger at low energies, and for net-kaon and net-charge the suppression depend weakly on energy at RHIC energies. The STAR measured monotonically decreasing net-charge $\sigma^2/M$ and increasing $S\sigma$ with increasing pseudo-rapidity coverage at $\sqrt{s_{NN}}=14.5$ GeV are quite consistent with the effects caused by resonance decays.

\begin{acknowledgments}
The author acknowledges the support of the FANEDD of PR China under Project No. 200523
and the NNSFC under Project Nos. 10305004, 11221504.
\end{acknowledgments}

%
% ****** End of file .tex ******

\end{document}